\documentstyle[12pt,epsfig]{article}

\newcommand{\re}{\mathop{\mathrm{Re}}\nolimits}
\newcommand{\im}{\mathop{\mathrm{Im}}\nolimits}

\catcode`@=11
\newcount\@tempcntc
\def\@citex[#1]#2{\if@filesw\immediate\write\@auxout{\string\citation{#2}}\fi
  \@tempcnta\z@\@tempcntb\m@ne\def\@citea{}\@cite{\@for\@citeb:=#2\do
    {\@ifundefined
       {b@\@citeb}{\@citeo\@tempcntb\m@ne\@citea\def\@citea{,}{\bf
?}\@warning
       {Citation `\@citeb' on page \thepage \space undefined}}%
    {\setbox\z@\hbox{\global\@tempcntc0\csname b@\@citeb\endcsname\relax}%
     \ifnum\@tempcntc=\z@ \@citeo\@tempcntb\m@ne
       \@citea\def\@citea{,}\hbox{\csname b@\@citeb\endcsname}%
     \else
      \advance\@tempcntb\@ne
      \ifnum\@tempcntb=\@tempcntc
      \else\advance\@tempcntb\m@ne\@citeo
      \@tempcnta\@tempcntc\@tempcntb\@tempcntc\fi\fi}}\@citeo}{#1}}
\def\@citeo{\ifnum\@tempcnta>\@tempcntb\else\@citea\def\@citea{,}%
  \ifnum\@tempcnta=\@tempcntb\the\@tempcnta\else
   {\advance\@tempcnta\@ne\ifnum\@tempcnta=\@tempcntb \else
\def\@citea{--}\fi
    \advance\@tempcnta\m@ne\the\@tempcnta\@citea\the\@tempcntb}\fi\fi}
\catcode`@=12
\begin{document}
\title{\vskip-3cm{\baselineskip14pt
\centerline{\normalsize DESY 01-146\hfill ISSN 0418-9833}
\centerline{\normalsize hep-ph/0110296\hfill}
\centerline{\normalsize October 2001\hfill}}
\vskip1.5cm
On the Field Renormalization Constant for Unstable Particles}
\author{{\sc Bernd A. Kniehl and Alberto Sirlin\thanks{%
Permanent address: Department of Physics, New York University,
4 Washington Place, New York, New York 10003, USA}}\\
{\normalsize II. Institut f\"ur Theoretische Physik, Universit\"at Hamburg,}\\
{\normalsize Luruper Chaussee 149, 22761 Hamburg, Germany}}

\date{}

\maketitle

\thispagestyle{empty}

\begin{abstract}
A recently proposed normalization condition for the imaginary part of the
self-energy of an unstable particle is shown to lead to a closed expression
for the field renormalization constant $Z$.
In turn, the exact expression for $Z$ is necessary, in some important cases,
in order to avoid power-like infrared divergences in high orders of
perturbation theory.
In the same examples, the width plays the r\^ole of an infrared cutoff and,
consequently, $Z$ is not an analytic function of the coupling constant.

\medskip

\noindent
PACS numbers: 11.10.Gh, 11.15.Bt
\end{abstract}

\newpage

The unrenormalized transverse propagator of a gauge boson is of the form:
\begin{equation}
{\cal D}_{\mu\nu}^{(u)}(s)=\frac{-iQ_{\mu\nu}}{s-M_0^2-A(s)},
\label{eq:unr}
\end{equation}
where $Q_{\mu\nu}=g_{\mu\nu}-q_\mu q_\nu/s$, $q_\mu$ is the four-momentum,
$s=q^2$, $M_0$ is the bare mass, and $A(s)$ is the unrenormalized self-energy.
An analogous expression holds for a scalar boson, with $-iQ_{\mu\nu}\to i$.
The complex position of the propagator's pole is given by
\begin{equation}
\bar s=M_0^2+A(\bar s).
\label{eq:sba}
\end{equation}
Combining Eqs.~(\ref{eq:unr}) and (\ref{eq:sba}), we have
\begin{equation}
{\cal D}_{\mu\nu}^{(u)}(s)=\frac{-iQ_{\mu\nu}}{s-\bar s-[A(s)-A(\bar s)]}.
\label{eq:dmn}
\end{equation}

Parameterizing $\bar s=m_2^2-im_2\Gamma_2$, where we employ the notation of
Ref.~\cite{sir}, and considering the real and imaginary parts of
Eq.~(\ref{eq:sba}), we see that
\begin{eqnarray}
m_2^2&=&M_0^2+\re A(\bar s),
\label{eq:m}\\
m_2\Gamma_2&=&-\im A(\bar s).
\label{eq:g}
\end{eqnarray}
If $m_2$ is identified with the renormalized mass, Eq.~(\ref{eq:m}) tells us
that the mass counterterm is given by $\delta m_2^2=\re A(\bar s)$.
This is to be contrasted with the conventional mass renormalization
\begin{equation}
M^2=M_0^2+\re A(M^2),
\end{equation}
where $M$ is the on-shell mass.
The great theoretical advantage of using $m_2$ and $\Gamma_2$ as the basis to
define mass and width is that they are intrinsically gauge-independent
quantities, while $M$ is known to be gauge dependent in
next-to-next-to-leading order \cite{sir,wil}.

The renormalized propagator ${\cal D}_{\mu\nu}^{(r)}(s)$ is obtained by
dividing Eq.~(\ref{eq:unr}) by the field renormalization constant
$Z=1-\delta Z$.
Recalling Eq.~(\ref{eq:m}), one readily obtains
\begin{equation}
{\cal D}_{\mu\nu}^{(r)}(s)=\frac{-iQ_{\mu\nu}}{s-m_2^2-{\cal S}(s)
+\re{\cal S}(\bar s)-\delta Z\left(s-m_2^2\right)},
\label{eq:dr}
\end{equation}
where
\begin{equation}
{\cal S}(s)\equiv ZA(s).
\label{eq:sza}
\end{equation}
Thus, the renormalized self-energy is given by
\begin{equation}
{\cal S}^{(r)}(s)={\cal S}(s)-\re {\cal S}(\bar s)
+\delta Z\left(s-m_2^2\right),
\end{equation}
where the second and third terms are identified with the mass renormalization
parameter,
\begin{equation}
\delta M^2\equiv Z\delta m_2^2=\re{\cal S}(\bar s),
\label{eq:dms}
\end{equation}
and the field renormalization counterterm, respectively.
Since the hermiticity of the Lagrangian density requires these counterterms to 
be real, $\delta Z$ must be chosen in such a way that, for real $s$,
$\re{\cal S}^{(r)}(s)$ is ultraviolet convergent to all orders of perturbation
theory.
Once this is done, $\im{\cal S}^{(r)}(s)=\im{\cal S}(s)=Z\im A(s)$ must also 
be ultraviolet convergent to all orders, since there are no further
counterterms available.
This means that $Z$ can be defined by choosing a suitable normalization
condition on $\im{\cal S}(s)=Z\im A(s)$.

Recently, a novel normalization condition for $\im{\cal S}(s)=Z\im A(s)$,
namely
\begin{equation}
\im{\cal S}\left(m_2^2\right)\equiv Z\im A\left(m_2^2\right)=-m_2\Gamma_2,
\label{eq:pal}
\end{equation}
was proposed independently in Eqs.~(22) and (23) of Ref.~\cite{pal} and in 
Ref.~\cite{nek}. 
The above relation between the width and the self-energy is known to be valid
at the one-loop order.
Eq.~(\ref{eq:pal}) extends its validity as an exact normalization condition,
valid to all orders.
While the objective of Ref.~\cite{pal} was to solve the notorious problem of
threshold singularities in the conventional definition of $Z$ \cite{fle}, that
of Ref.~\cite{nek} was to provide a second normalization condition for the
systematic order-by-order removal of ultraviolet divergences in
${\cal S}^{(r)}(s)$.

An interesting feature of the analysis of Ref.~\cite{pal} is that it leads to
exact, closed expressions for $Z$.
This may be understood immediately by combining Eq.~(\ref{eq:pal}) with
Eqs.~(\ref{eq:g}) and (\ref{eq:sza}):
\begin{equation}
\im{\cal S}\left(m_2^2\right)=\im A(\bar s)=\frac{\im{\cal S}(\bar s)}{Z}.
\end{equation}
Thus,
\begin{equation}
Z=\frac{\im{\cal S}(\bar s)}{\im{\cal S}\left(m_2^2\right)}
=1-\frac{\im\left[{\cal S}(\bar s)-{\cal S}\left(m_2^2\right)\right]}
{m_2\Gamma_2},
\label{eq:z}
\end{equation}
where we again employed Eq.~(\ref{eq:pal}).
Eq.~(\ref{eq:z}) tells us that, once the normalization condition of
Eq.~(\ref{eq:pal}) is adopted, the field renormalization counterterm is given
by the closed expression
\begin{equation}
\delta Z=\frac{\im\left[{\cal S}(\bar s)-{\cal S}\left(m_2^2\right)\right]}
{m_2\Gamma_2}.
\label{eq:dz}
\end{equation}
In particular, the renormalized self-energy my be written as
\begin{equation}
{\cal S}^{(r)}(s)={\cal S}(s)-\re{\cal S}(\bar s)
+\frac{\im\left[{\cal S}(\bar s)-{\cal S}\left(m_2^2\right)\right]}
{m_2\Gamma_2}\left(s-m_2^2\right).
\end{equation}
Combining Eq.~(\ref{eq:z}) with Eq.~(\ref{eq:sza}), $Z$ may be also expressed
in terms of the unrenormalized self-energy, as
\begin{equation}
Z=\frac{1}{1+\im\left[A(\bar s)-A\left(m_2^2\right)\right]/(m_2\Gamma_2)},
\label{eq:zpa}
\end{equation}
the expression given in Eq.~(23) of Ref.~\cite{pal}.
As explained in that paper, Eq.~(\ref{eq:zpa}), in conjunction with
Eq.~(\ref{eq:g}), leads to $Z\im A\left(m_2^2\right)=-m_2\Gamma_2$ as a
mathematical identity, a result that coincides with the normalization 
condition of Eq.~(\ref{eq:pal}).

In general, $\Gamma_2={\cal O}(g^2)$, where $g$ is a generic gauge coupling.
If $\delta M^2=\re{\cal S}(\bar s)$ and
$\delta Z=\im\left[{\cal S}(\bar s)-{\cal S}\left(m_2^2\right)\right]
/(m_2\Gamma_2)$ admit expansions in powers of $\Gamma_2$, one readily obtains
the expressions for the counterterms to all orders.
For instance,
\begin{eqnarray}
\delta M^2&=&R-II^\prime-\frac{I^2}{2}R^{\prime\prime}+\frac{I^3}{6}I^{(3)}
+\frac{I^4}{24}R^{(4)}+\cdots
\nonumber\\
&=&\sum_{n=0}^\infty(-1)^n\left[\frac{I^{2n}}{(2n)!}R^{(2n)}
-\frac{I^{2n+1}}{(2n+1)!}I^{(2n+1)}\right],
\nonumber\\
\delta Z&=&-R^\prime+\frac{I}{2}I^{\prime\prime}+\frac{I^2}{6}R^{(3)}
-\frac{I^3}{24}I^{(4)}+\cdots
\nonumber\\
&=&\sum_{n=0}^\infty(-1)^{n+1}\left[\frac{I^{2n}}{(2n+1)!}R^{(2n+1)}
-\frac{I^{2n+1}}{(2n+2)!}I^{(2n+2)}\right],
\label{eq:exp}
\end{eqnarray}
where $R\equiv\re{\cal S}(s)$, $I\equiv\im{\cal S}(s)$, the primes and 
superscripts $(n)$ indicate derivatives with respect to $s$, and all the
functions are evaluated at $s=m_2^2$.
Separating out the contributions of $i$-th loop order ($i=1,2,\ldots$) and
taking into account differences in the sign conventions, the first few terms
of Eq.~(\ref{eq:exp}) coincide with the results obtained in Ref.~\cite{nek} by
considering the systematic order-by-order renormalization.
The leading terms in Eq.~(\ref{eq:exp}) are the results of the conventional
analysis valid in the narrow-width approximation, except that $R$ and 
$R^\prime$ are evaluated at $m_2^2$, rather than $M^2$.
The other terms in Eq.~(\ref{eq:exp}) represent further contributions beyond
that approximation.

In some important cases, however, the expansions of Eq.~(\ref{eq:exp}) are
ill-defined, since they lead to power-like infrared divergences.
Examples include contributions to the $W$-boson and unstable-quark
self-energies involving $(W,\gamma)$ and $(q,g)$ virtual contributions,
respectively, where $\gamma$, $g$, and $q$ denote photons, gluons, and
unstable quarks.
For instance, it is well known that the one-loop $(W,\gamma)$ virtual
contribution to the unrenormalized $W$-boson self-energy $A(s)$ contains a
term $c\left(s-m_2^2\right)\ln\left[\left(m_2^2-s\right)/m_2^2\right]$, where
$c=(\alpha/2\pi)(\xi_\gamma-3)$ and $\xi_\gamma$ is the gauge parameter
associated with the photon \cite{pas}.
This leads to a logarithmic singularity in $R^\prime$, namely
$c\lim_{s\to m_2^2}\ln\left[\left(m_2^2-s\right)/m_2^2\right]$, which, as is
well known, is conventionally regularized with an infinitesimal photon mass or
by means of dimensional regularization \cite{mar}.
However, higher order derivatives such as $R^{(n)}$ develop singularities
proportional to $c\lim_{s\to m_2^2}\left(m_2^2-s\right)^{1-n}$ ($n\ge2$),
i.e.\ power-like infrared singularities of any order.
A similar behaviour arises from the one-loop $(q,g)$ virtual contribution to
the unstable-quark self-energy.
Clearly, such a catastrophic behaviour is a strong indication that, in such
cases, the expansions in Eq.~(\ref{eq:exp}) are ill-defined and highly
unphysical.

In considering the $(W,\gamma)$ virtual contribution in the resonance region
$s\approx m_2^2$, it is important to take into account the effect of
self-energy insertions in the $W$-boson line.
As explained in Refs.~\cite{pas,wac}, this leads to the replacement
\begin{equation}
c\left(s-m_2^2\right)\ln\frac{m_2^2-s}{m_2^2}
\to c(s-\bar s)\ln\frac{\bar s-s}{\bar s}.
\label{eq:rep}
\end{equation}
If the modified form of Eq.~(\ref{eq:rep}) is inserted in Eq.~(\ref{eq:exp}),
the power-like infrared divergences are avoided, but terms of the form
$c(-im_2\Gamma_2)^{1-n}$ ($n\ge2$) are generated, which signal a breakdown of
the perturbative expansion.
Thus, inverse powers of $\Gamma_2$ would occur in $R^{(n)}$ and $I^{(n)}$ for
$n=3,5,\ldots$ and $n=2,4,\ldots$, respectively.

These catastrophic problems can be neatly avoided by employing the exact,
closed expressions of Eqs.~(\ref{eq:dms}) and (\ref{eq:dz}).
For instance, $c(s-\bar s)\ln[(\bar s-s)/\bar s]$ contributes zero to
$\delta M^2$ [cf.\ Eq.~(\ref{eq:dms})] and $-c\ln a$, where
$a=\Gamma_2/\sqrt{m_2^2+\Gamma_2^2}$, to $\delta Z$ [cf.\ Eq.~(\ref{eq:dz})].
Thus, in this contribution, the width plays the r\^ole of an infrared cutoff.
Since generally $\Gamma_2={\cal O}(g^2)$, we also see that $\delta Z$ is not
an analytic function of $g^2$ in the neighborhood of $g^2=0$.

An expression for the renormalized propagator, alternative to
Eq.~(\ref{eq:dr}), is obtained by dividing Eq.~(\ref{eq:dmn}) by $Z$:
\begin{equation}
{\cal D}_{\mu\nu}^{(r)}(s)=\frac{-iQ_{\mu\nu}}
{s-\bar s-[{\cal S}(s)-{\cal S}(\bar s)]-\delta Z(s-\bar s)}.
\label{eq:dtw}
\end{equation}
Writing \cite{pas}
\begin{equation}
\frac{\bar s-s}{\bar s}=\rho(s){\rm e}^{i\theta(s)},
\label{eq:rho}
\end{equation}
where
\begin{eqnarray}
\rho(s)&=&\frac{1}{m_2}\sqrt{\frac{\left(s-m_2^2\right)^2+m_2^2\Gamma_2^2}
{m_2^2+\Gamma_2^2}},
\nonumber\\
\rho(s)\sin\theta(s)&=&-\frac{s\Gamma_2}{m_2\left(m_2^2+\Gamma_2^2\right)}
\qquad(-\pi\le\theta\le\pi),
\end{eqnarray}
and using Eq.~(\ref{eq:dz}), the contribution of
$c(s-\bar s)\ln[(\bar s-s)/\bar s]$ to Eq.~(\ref{eq:dtw}) becomes
\begin{equation}
\frac{-iQ_{\mu\nu}}{(s-\bar s)\{1-c[\ln \rho(s)-\ln a+i\theta(s)]\}}.
\label{eq:qmn}
\end{equation}
At $s=m_2^2$, we have $\rho\left(m_2^2\right)=a$ and 
$\sin\theta\left(m_2^2\right)=-m_2/\sqrt{m_2^2+\Gamma_2^2}$, so that the
expression between curly brackets in Eq.~(\ref{eq:qmn}) is purely imaginary.
Furthermore, $\theta\left(m_2^2\right)\approx-\pi/2$ in the case
$\Gamma_2\ll m_2$.
Thus, with the choice of Eq.~(\ref{eq:dz}), the r\^ole of $\delta Z$ is to
cancel the contribution $\ln\rho(s)$ at $s=m_2^2$, which depends
logarithmically on $\Gamma_2$.
These results may be understood directly by setting $s=m_2^2$ in
Eqs.~(\ref{eq:dr}) or (\ref{eq:dtw}) and recalling Eqs.~(\ref{eq:pal}) and
(\ref{eq:rho}).
On the other hand, far away from the resonance region, i.e.\ for
$\left|s-m_2^2\right|\gg m_2\Gamma_2$, $\ln\rho(s)$ does not depend
logarithmically on $\Gamma_2$, while $\ln[\rho(s)/a]$ does.
The latter feature is not surprising, since away from the resonance region
${\cal S}(s)$ has a regular behaviour, while $\delta Z$ is logarithmically
divergent in the limit $\Gamma_2\to0$.

As mentioned before, the original motivation that led to Eq.~(\ref{eq:zpa}) or
its equivalent, Eq.~(\ref{eq:z}), was to solve the problem of threshold
singularities in the evaluation of $Z$, which occurs when the mass of the
unstable particle is degenerate with the sum of masses of a pair of
contributing virtual particles \cite{pal}.
Since $m_\gamma=m_g=0$, the contributions of the $(W,\gamma)$ and $(q,g)$
virtual pairs to the $Z$ factors of the $W$ boson and the unstable quark $q$
are particular, albeit very important, examples of threshold singularities.
It is for this reason that the definition of the $Z$ factor, embodied in the
exact, closed expressions of Eqs~(\ref{eq:z}) and (\ref{eq:zpa}), provides a
consistent formulation to treat the associated ``infrared contributions."

In summary, we have shown that the normalization condition of
Eq.~(\ref{eq:pal}) leads to exact, closed formulae for the field
renormalization constant of the unstable particle [Eq.~(\ref{eq:z}) or its
equivalent, Eq.~(\ref{eq:zpa})] and that, in some important cases, these
expressions play an important r\^ole in the evaluation of $Z$ beyond the
narrow-width approximation.

\vspace{1cm}
\begin{center}
{\bf Acknowledgements}
\end{center}
\smallskip

A.S. is grateful to the Theory Group of the 2$^{\rm nd}$ Institute for
Theoretical Physics for the hospitality extended to him during a visit when
this manuscript was prepared.
The work of B.A.K. was supported in part by the Deutsche
Forschungsgemeinschaft through Grant No.\ KN~365/1-1, by the
Bundesministerium f\"ur Bildung und Forschung through Grant No.\ 05~HT1GUA/4,
and by the European Commission through the Research Training Network
{\it Quantum Chromodynamics and the Deep Structure of Elementary Particles}
under Contract No.\ ERBFMRX-CT98-0194.
The work of A.S. was supported in part by the Alexander von Humboldt
Foundation through Research Award No.\ IV~USA~1051120~USS, and by the
National Science Foundation through Grant No.\ PHY-0070787.

\end{document}